\documentclass[nonacm, acmlarge,screen]{acmart}

\AtBeginDocument{%
  \providecommand\BibTeX{{%
    \normalfont B\kern-0.5em{\scshape i\kern-0.25em b}\kern-0.8em\TeX}}}

\copyrightyear{2023}
\acmYear{2023}
\setcopyright{cc}
\setcctype{by-nc-sa}
\acmConference[CHI '23 Workshop]{CHI '23 Designing Technology and Policy Simultaneously Workshop}{April 23,
2023}{Hamburg, Germany}
\acmDOI{}
\acmISBN{}

\newcommand\citetodo[1]{\textcolor{olive}{[CITE]}}

\usepackage{enumitem}
\usepackage{xspace}
\newcommand{\newmethodlong}[0]{integrative City-AV-Policy Simulation\xspace}
\newcommand{\newmethod}[0]{iCAPS\xspace}

\begin{document}
\fancyhead[RO]{\fontfamily{LinuxBiolinumT-TLF}\fontsize{8}{10}\selectfont CHI '23 Workshop, April 23, 2023, Hamburg, Germany}

\fancyhead[LE]{\fontfamily{LinuxBiolinumT-TLF}\fontsize{8}{10}\selectfont Designing Technology and Policy Simultaneously Workshop}

\title[Towards Prototyping Driverless Vehicles, Cities, and Policies Simultaneously]{Towards Prototyping Driverless Vehicle Behaviors, \\City Design, and Policies Simultaneously}



\author{Hauke Sandhaus}
\email{hgs52@cornell.edu}
\orcid{0000-0002-4169-0197}
\affiliation{%
  \institution{Cornell University, Cornell Tech}
  \streetaddress{2 West Loop Rd}
  \city{New York}
  \state{New York}
  \country{USA}
  \postcode{10044}
}

\author{Wendy Ju}
\orcid{0000-0002-3119-611X}
\email{wendyju@cornell.edu}
\affiliation{%
  \institution{ Cornell Tech}
  \streetaddress{2 West Loop Rd}
  \city{New York}
  \state{New York}
  \country{USA}
  \postcode{10044}
}

\author{Qian Yang}
\orcid{0000-0002-3548-2535}
\email{qianyang@cornell.edu}
\affiliation{%
  \institution{Cornell University}
  \city{Ithaca}
  \state{New York}
  \country{USA}
  \postcode{14853}
}

\renewcommand{\shortauthors}{Sandhaus, Ju, and Yang}

\maketitle

  \vskip -0.2em%

\section{Introduction}
Driverless vehicles (commonly referred to as Autonomous Vehicles, AVs) promise to improve cities in ways that would not have been possible otherwise~\cite{World_Economic_Forum2020-ox,Appleyard2014-ph, The_Intelligent_Transportation_Society_of_America2021-ew}. 
AVs can reduce crashes, deaths, and injuries~\cite{Golbabaei2020TheRO,PETTIGREW201964,fagnant2015preparing}. 
They can allow easier travel for the handicapped, the elderly, and those living in poorly-connected neighborhoods~\cite{Gilbert-Sociotechnical}. 
They can decrease vehicle emissions by reducing congestion and cruising-for-parking~\cite{Rojas-Rueda2017-xl}. 
These promises of safer, more equal, more livable cities have been key motivations for AV innovation.

Yet driverless vehicles' societal benefits are far from guaranteed; Realizing them requires thoughtful AV driving algorithm design as well as city design and policy innovations~\cite{Faisal2019UnderstandingAV, Acheampong2021-ql, Metz2018-jn,Yigitcanlar2019DisruptiveIO, Golbabaei2020TheRO, Meeder2017AutonomousVP, Staricco2019-gn, PETTIGREW201964, Bobisse2019-wj, Gonzalez-Gonzalez2019-cn, Rouse2021-ph, Legacy2019-me,Bosch2018-pl, Merlin2019-fb, Marx2022-rm}.
Just making AVs available to the public alone involves complex policy questions (e.g., liability, privacy, insurance)~\cite{fagnant2015preparing} and potential city-scale infrastructural changes (e.g., should cities create new dedicated lanes for autonomous food-delivery robots?~\cite{food-delivery-robot-bikelane-wired}); Much more so in fulfilling the ambitious goal of better AV-infused cities. 

But how? 
The decisions around AV, city, and policy design are not just related. They form a difficult ``\textit{knot}''~\cite{Gilbert-Sociotechnical,Wu2018-oy, Brozynski2022-yv}:
Changes to any one of the three---AVs' driving pattern, road geometry, or traffic law---must be compatible with the other two to work; 
Meanwhile, changes to any of them also change the other two.
Taking a lesson from earlier HCI work~\cite{policy-knot, buchanan1992wicked}, we argue that the path towards better AV cities is \textit{not} to match city designs and policies with AVs' technological innovations, but an iterative prototyping process: Innovations happen step-wise as the ``\textit{knot}'' of AV, city, and policy design loosens and tightens, unwinds and reties.

\textbf{This paper takes the knot-like entanglement among AV, city, and policy design as a starting point and asks: How can innovators \textit{innovate} AVs, urban design, and policies simultaneously but productively toward better AV cities?}
The paper has two parts.
First, we trace how the ``\textit{knot}'' is tied. We map out the interconnections among the many AV, city, and policy design decisions, based on a cross-disciplinary literature review covering human-computer/robot interaction (HCI/HRI), transportation science, urban studies, law and policy, operations research, economy, and philosophy.
This map can help innovators identify design constraints and opportunities across the traditional AV/city/policy design disciplinary bounds.

Next, we investigate how innovators might simultaneously prototype AV, city, and policy innovations. 
We review the respective methods for AV, city, and policy design, and then identify key barriers in combining them for simultaneous prototyping: (1) Organizational barriers to AV-city-policy design collaboration, (2) computational barriers to multi-granularity AV-city-policy simulation, and (3) different assumptions and goals in joint AV-city-policy optimization.
We discuss two broad approaches that can potentially address these challenges, namely ``\textit{low-fidelity \newmethodlong}'' (\newmethod) and ``\textit{participatory design optimization}''.
Taken together, this work aspires to catalyze a more principled discussion around how AV designers, urban planners, and policymakers can more productively work together to innovate AVs and better cities.



\section{Mapping AV, City, and Policy Design Issues}\label{sec:rw}


\begin{figure*}[h]
    \includegraphics[width=0.9\textwidth]{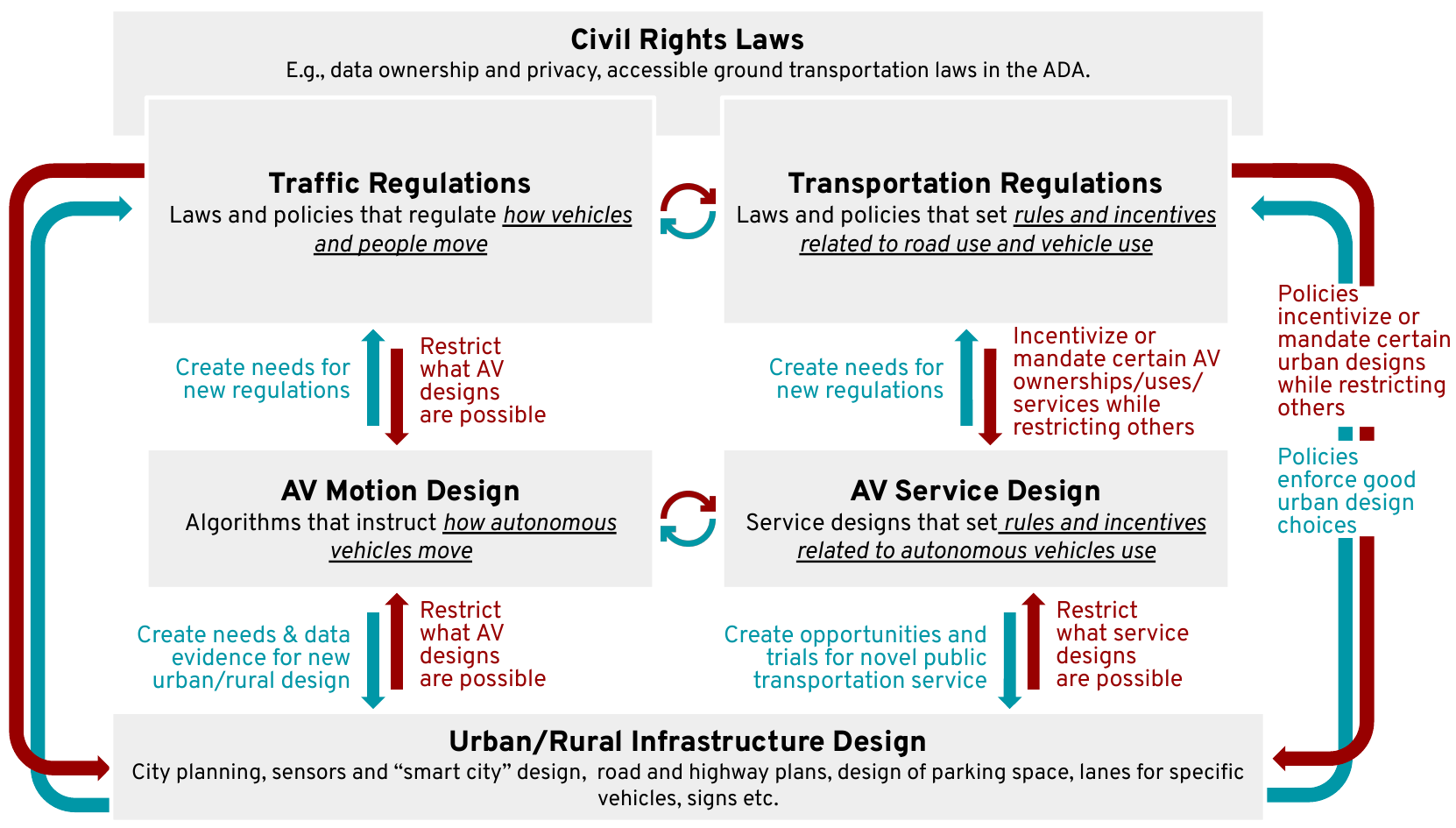}

    \captionsetup{width=0.9\textwidth}
  \caption{Mapping the many ways driverless vehicles' behaviors, service design, urban design, and law and policy influence each other. They offer each other both design opportunities (green arrows) and constraints (red arrows).}
  \Description{A diagram mapping the many ways driverless vehicles' motion design, service design, urban design, and law and policy influence each other. They offer each other both new opportunities (green arrows) and constraints (red arrows).}
\label{fig:interactions}
\end{figure*}  



In this chapter, we trace how the AV-city-policy ``\textit{knot}'' is tied~(Figure~\ref{fig:interactions}.) 

From top to bottom: \textit{Civil rights laws and policies} set rules that apply to all players in transportation. For example, they require vehicle companies and vehicle designs to allow free sharing of car-repair-related data~\cite{Robertson2020-uq}; they require cities to offer a certain amount of accessible parking and ride-sharing services to provide a certain amount of accessible vehicles~\cite{Reck2018-sq, ADA_National_Network2018-bq}. 
\textit{Traffic regulations} instruct how vehicles and people should move (e.g., pedestrians law, school bus stopping law, etc.) 
\textit{Transportation regulations} set rules and incentives related to road use and vehicle use (e.g., authorizing protected planes for bikes and scooters~\cite{Pierce2022-vq}.) 
Within these legal and regulatory bounds, \textit{AV motion designers} design algorithms that instruct how AVs move in relation to city infrastructure and other road users; \textit{AV service designers} create additional rules and incentives and shape AVs' and service users' behaviors (e.g., via dynamic pricing~\cite{Kaddoura2020-vw}).
If proven effective, good vehicle designs can become policy mandates (e.g., the invention of seat belts) and can compel cities to adopt new accommodating infrastructure (e.g.,  Uber's popularity led to dedicated ride-hailing boarding zones at airports; Uber is even becoming a part of public transit in some cities~\cite{hawkins_2021}).   

This map reaffirms that realizing AVs' societal benefits \textit{requires} coordination among AV, city, and policy design decisions~\cite{Matowicki2020-vy,Aoyama2021-jj,Duarte2018-wg,OToole2020-so,Freemark2020-ms,Staricco2019-gn}. 
Empirical evidence from small-scale AV trials and large-scale traffic simulations agree~\cite{Soteropoulos2019-tw,Sperling2018-vk, Freemark2020-ms, Nacto2019-pe,Riggs2019-lx, Niroumand2023-an,Staricco2019-gn,Bosch2018-pl, Merlin2019-fb, Marx2022-rm}.
For example, without AV services, public parking, and regulations that incentivize shared or publicly owned AVs, AVs are likely to increase congestion, regardless of how the AVs' navigation algorithms are optimized to reduce congestion~\cite{Campbell2018-rr, Sanguinetti2019-de, Kaddoura2020-vw}.
Similarly, making future cities more accessible to people with mobility disabilities needs more than accessible AVs.
Incorporating accessible AVs into public transport services can radically scale up AV's accessibility benefits, while corresponding laws and policies could help minimize risks (e.g., by mandating blind driver boarding zones or way-finding beacons)~\cite{Reck2018-sq, ADA_National_Network2018-bq,Ustwo2015-ck}.


\section{Exploring How to Design AVs, Cities, and Policies Simultaneously}\label{sec:proposition}

The previous chapter has argued that AV, city, and policy design issues are inherently entangled and that innovators should consider them simultaneously. But how? In this chapter, we first demonstrate the lack of methods and tools for prototyping AV, city, and policy innovations simultaneously. We then identify promising paths forward in developing such new methods.

\subsection{Respective Methods of Designing AVs, Cities, and Policies}
AV, city, and policy innovators today share few common design methods~\cite{Rouse2021-ph,Yang2019-nr}.~For example:

\begin{itemize}[leftmargin=*, itemsep=2pt]
    \item \textit{AV driving behavior design uses road trials and microscopic driving simulations.~}AV designers train machine learning models to ``\textit{drive}'' driverless vehicles, typically using datasets that capture how people drive on today's roads or test roads~\cite{Pettersson2017-wi,Baltodano2015-yf,Rothenbucher2016-ja,Dowling2022-wu,Suo2021-vm}. Recently, AV designers and researchers have also become interested in immersive driving simulation studies~\cite{Dosovitskiy2017-jb,NVIDIA_Corporation2023-tn, Amini2021-qu, Jin2018-sw, Craighead2007-vp,driving-like-you}. 
Unlike road trial datasets that capture drivers' naturalistic behaviors in messy real-world driving conditions and interactions, these simulation studies capture such behaviors in simple, controlled road situations (e.g., two cars merging on a highway) typically with only 1-2 vehicles and no pedestrians~\cite{Osorio2011-kj, Kavas-Torris2021-xm, Wang2017-qe,Yang2021-oz, Pell2017-al}. 
With few rare exceptions~\cite{Goedicke2022-ys}, such simulations typically do not consider regional differences in city environment or driving style. 
Further, no literature to our knowledge has reported whether and how AV designers have incorporated these insights into AVs' driving behavior models.

    \item \textit{City planning uses design patterns and macroscopic traffic simulation.~}Urban planners heavily rely upon established design patterns~\cite{alexander1977pattern}. They choose and modify the patterns using maps, geographic information systems, or street cross-profiles~\cite{Riggs2017-hm,Lee2022-vd, Remix_by_Via2021-vm,Ziliaskopoulos2000-az, Rasheed2020-op,Charitonidou2022-zg}.
Additionally, urban planners~\cite{Zhang2022-uh, Dembski2020-gk} and policy makers~\cite{White2021-vo, Charitonidou2022-zg} use computational simulations to identify optimal city-level designs, such as zoning requirements for loading areas that offer the maximum convenience and safety and minimal traffic disruption. 

    \item \textit{Street-level city design and policy-making use design patterns and participatory design.~}Street-level road geometry and infrastructure design also relies on established design patterns and guidelines, for example, in designing bus lane arrangement that causes minimal traffic disruption~\cite{Nacto2013-zd,Federal_Highway_Administration2023-wu}. In recent years, participatory design (PD) has gained popularity, incorporating citizen inputs into the process of neighborhood and city design~\cite{Streetplans_Collaborative2016-xt, Lydon2015-nn, Pfeifer2013-ew}.

    \item \textit{Traffic and transportation law-making inherits prevailing law-making processes and sometimes adds quantitative modeling, regulatory sandboxes, and more.~}AV regulation is an active research topic in legal fields. Its approaches largely echo traditional traffic law-making approaches. That said, the last couple of years have witnessed new approaches emerging. For example, conventional transportation research uses mathematical modeling to address policy questions (e.g., how tolls and fines impact traffic flows)~\cite{Krajzewicz2012-ah}. Yet, this approach assumes a top-down execution of proposed policies and can appear detached from the messy policy-making cycles and the rapid AV technological advances in practice~\cite{Marsden2017-lu}. In response, lawmakers have adopted participatory approaches in making AV policies~\cite{van2022politics,lyons2022driverless}. They have also created ``\textit{regulatory sandbox}'' for AV safety regulations as a means of providing a flexible regulatory environment for companies to iteratively test emerging technologies and business models~\cite{oecd2020-regulatory-sandbox,burd2021regulatory, BMWK-Federal_Ministry_for_Economics_Affairs_and_Climate_Action2022-ck}. 
\end{itemize}

\begin{figure*}[t]
      \centering
    \includegraphics[width=0.9\textwidth]{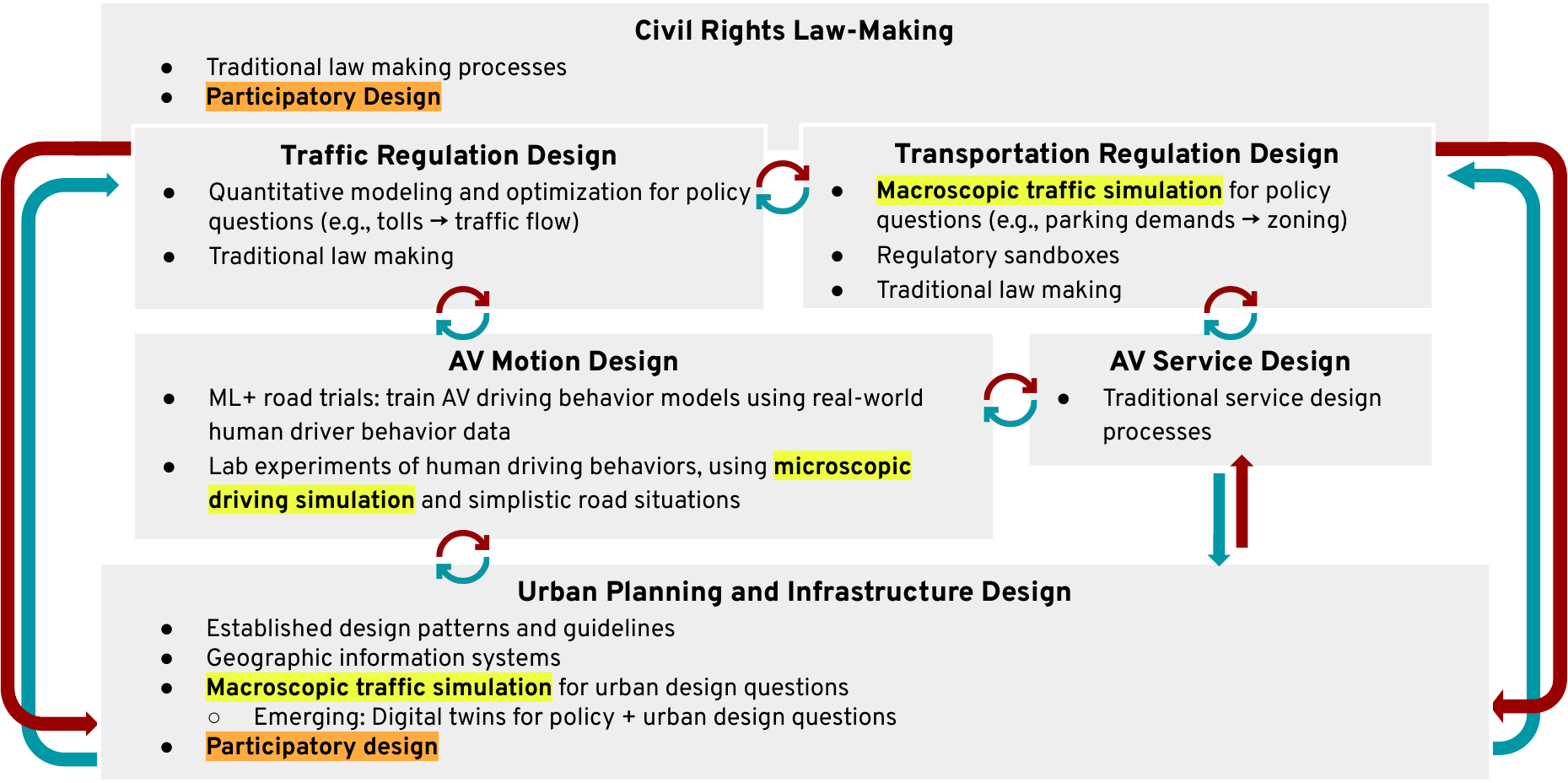}
\captionsetup{width=0.6\textwidth}    
\caption{Common methods for prototyping AV, city design, and policy innovations. Shared methods (highlighted in yellow and orange) are few.}
      \Description{Common methods for prototyping AV, city design, and policy innovations. }
    \label{fig:methods}
\end{figure*}

All but two approaches can arguably support multiple tasks across AV, city, and policy design.
First, traffic and city simulations have the potential to support simultaneous urban infrastructure and policy design (Figure~\ref{fig:methods} yellow highlights)~\cite{Marcucci2020-qg,White2021-vo,Riggs2020-yc}. 
Beyond traditional simulation techniques, the emergent ``\textit{digital twins}'' can appear particular promising~\cite{shahat2021city}. Digital twin refers to a dynamic 3D model that both mirrors a city's physical form and contains the city's information throughout its life cycle (e.g., street network space syntax, commuting patterns, the socioeconomic data of a neighborhood).
Digital twins' rich data can allow for machine learning models that can answer ``\textit{what if}'' policy questions~\cite{nochta2021socio}.
It also offers a shared platform, a rare boundary object, for various stakeholders to discuss city goals~\cite{hamalainen2020smart}.  
However, digital twin technologies and datasets are emergent; no research we are aware of has yet to implement them for coordinating AV-city-policy designs~\cite{Riggs2020-yc}.

Second, Participatory Design (PD) methods have been supporting policy making and urban planning, but not yet supporting AV behavior or service design~(Figure~\ref{fig:methods} orange highlights).  



\subsection{Challenges in Designing AV, City, and Policy Simultaneously}
So far, digital twin simulations and PD methods appear the most promising in enabling simultaneous prototyping of future AVs, cities, and policies.
However, our analysis shows several outstanding barriers that remain.

\vspace{3pt}\noindent
\textbf{Different organizations own the design decisions around AVs, different components of city infrastructure, and different regulations.~}
Let us start with the biggest elephant in the room: that different private corporations and government agencies own the various AV, city, and policy design decisions. 
Designing them simultaneously requires designers across these agencies to navigate the organizational barriers and collaborate. This is not an easy task. 
For example, while multiple AV designers argued that small tweaks to traffic lights and road signs could instantaneously improve AV model performance and safety~\cite{Niroumand2023-an,Gilbert-Sociotechnical,Riggs2019-lx, Nacto2019-pe}, they simply are not in the position to adopt these proposals.

Moreover, different driverless vehicles operate under different traffic and transportation regulations (e.g., AV scooters will likely be subject to micro-mobility laws, while low-haul trucks will be to Hours of Service regulations in the U.S.) They will likely drive subtly differently as companies design them without coordination. 
They will also likely achieve wide adoption at different times and use different parts of the city infrastructure. The wide variety of future AVs will add further complexities to the already complex AV-city-policy design landscape.

\vspace{3pt}\noindent
\textbf{Prototyping AV-human-city interactions both at individual and societal levels is computationally (too) demanding.~}
AV, city, and policy design operate at different scales.
Take simulation methods as an example. 
Designing AV driving behavior models requires microscopic simulations to prototype the interactions between individual vehicles and pedestrians.
Because ideally, these models can drive AVs on all roads and in all situations, these simulations typically focus on simple, abstracted, and isolated road situations (e.g., the merging of two cars~\cite{Saidallah2016-zx}.)
In contrast, urban planning and policy-making require macroscopic simulations to prototype traffic flows, their longitudinal patterns, and their holistic impact on the city~\cite{Gilbert-Sociotechnical}.
Theoretically, high-resolution simulations (e.g., digital twins) can depict both individual road encounters and holistic traffic patterns and impacts. However, this approach is computationally highly demanding.
One 2021 study applied this integrative simulation approach to designing charging locations for electric vehicles. It only compared two designs and their impact on charging demand~\cite{nochta2021socio}. It is unclear how feasible it is to simulate multiple, more complex AV/city/policy design proposals and investigate their interactions or what computational resources designers need to do so.  
\vspace{3pt}\noindent
\textbf{Different assumptions and design goals underlie different AV, city, and policy design efforts.~}
Even if AV designers, urban planners, and policymakers can simultaneously \textit{visualize} their designs in a shared simulation system, deciding how to \textit{improve}, coordinate, and make trade-offs between their designs entails additional complications.
Today, AV, city, and policy design efforts often operate under different assumptions and pursue different goals.
For example, urban planners and policymakers tend to (sometimes unknowingly) consider AVs as part of the public infrastructure and optimize their designs towards maximal public good (e.g., less congestion~\cite{Metz2018-jn}, easier road maintenance~\cite{Gilbert-Sociotechnical}, and less environmental impact~\cite{Felipe-Falgas2022-dz}).
In contrast, companies designed AVs toward maximum customer value (e.g., they tend to personalize, rather than coordinate, each AV's driving behaviors).
These varying assumptions led to insistent debates around how AVs should move, who has a say in how AVs should move, and how AVs should fit into the public sphere~\cite{Wadud2021-zs, Rojas-Rueda2017-xl, Campbell2018-rr}.

Participatory Design (PD) methods have long addressed such ``\textit{design goal}'' questions in the public sphere. PD helps urban planners and policymakers to elicit and incorporate local knowledge and citizen needs and values (Figure~\ref{fig:methods} orange highlights). However, PD can struggle with designing complex machine learning (ML) models that operate in the public sphere because citizens and stakeholders often struggle to understand how ML works or what it can do for them specifically~\cite{delgado2021stakeholder}. These challenges are likely to apply to the models that govern AV behaviors.

\section{Paths Toward Prototyping AV, City, and Policy Innovations Simultaneously}
We see a need for more research into the AV-city-policy ``\textit{knot}'' and new methods and tools that enable more productive design collaborations. HCI and its adjacent fields are well-positioned to address these challenges, given their focus on human-centered design, prototyping methods, and interdisciplinary work.
To jump-start this discussion, we identify three promising directions that merit further research.


\subsection{Collaboration Tools and Communities for AV Designers, Urban Planners, and Policymakers}

No single, existing organization has the expertise or authority to create all the AV technologies, policies, and urban designs that a more livable, equal, and sustainable future city requires.
Moreover, the knot-like interconnections among AV, city, and policy design issues require designers and agencies to collaborate closely throughout their design processes.
Therefore, cultivating productive collaborations and communities among AV designers, city planners, and policymakers is crucial.

Establishing intimate collaboration across AV companies and government agencies can appear daunting, but it is possible. To varying degrees, such vehicle-city-policy design collaboration already exists for ensuring car safety, including AV safety (exemplified by the aforementioned AV regulatory sandboxes.) Although not yet covering other AV issues, new collaborations are also emerging for designing smart cities~\cite{Kagerbauer2021-in, Kreienbaum2022-bc} and ride-sharing policies~\cite{T4a2020-jn, Stone2016-yl, Open_Mobility_Foudation2018-ck}.
We see an opportunity to harness and strengthen these collaborations for AV innovations broadly (e.g., for sustainability, social equality, and accessible mobility).
\subsection{Low-fidelity, \newmethodlong (\newmethod)} 
We promote the idea of creating low-fidelity simulation tools that can visualize and predict the impact of AV behavior, city, or policy design proposals on traffic and cities.
When zoomed in, the simulation shows designers how their design proposal changes the interactions among individual vehicles, pedestrians, road geometry, and infrastructure designs. Zoomed out, it visualizes the proposal's impacts on traffic flows, traffic patterns, and the city and city life at large.
Such a shared platform can allow synchronous prototyping and easy collaboration among AV, city, and policy designers.

Several proven technologies can serve as building blocks for \newmethod tools.
First is digital twins, which offer rich semantic data and high-fidelity visualizations (point clouds) of buildings, roads, traffic signs, and other infrastructure.
Second, making these point clouds modifiable can transform digital twins from a visualization tool to a street-level city design tool.
Finally, add machine learning (ML) models that approximate road user behaviors (for example, ML models that predict how a change in driverless cars' right-of-road policy might improve safety~\cite{Suo2021-vm}.)
Such ML models, combined with digital twin visualizations, can show the effects of AV designers' and policymakers' proposals, therefore are valuable prototyping tools.
Making \newmethod feasible also needs additional research advances. Central to these advances is the idea of low-fidelity prototyping: Visualizing a city, its AV behaviors, and its policies in high resolution is computationally demanding (if not impossible) for the simulation system and cognitively demanding for designers, not to mention running multiple ML models on top to predict the impact of various design proposals. Therefore, abstraction is necessary: The visualizations and ML predictions must be precise and realistic \textit{only when relevant to the design proposal in question}. ---How can we better understand the visualization and ML needs over the process of integrative AV-city-policy prototyping? What new data visualization techniques and road-user-behavior-simulation models can better serve these needs on-demand, within reasonable computational constraints? 
For researchers to investigate these questions, what policies and tools need to be in place first to allow easy, ethical, and inoperable data-sharing across AV companies, urban planning, and regulatory agencies? 
Future research that investigates these questions can play a critical role in enabling integrative AV-city-policy designs.

\subsection{Participatory AV-City-Policy Design Optimization}

Varying assumptions (e.g., AV ownership) and competing goals (e.g., safety versus efficiency) have been driving various AV, city, and policy innovation efforts, leading to sometimes incompatible designs that pull in different directions. 
Integrative simulation and Participatory Design (PD) methods can help. The former offers a shared platform for innovators to confront their assumption differences and explore alternatives. The latter helps align various road users' and stakeholders' goals and priorities through discussion and debate.
But how can these stakeholder discussions produce actionable insights for design? 

We propose a new workflow, namely ``\textit{participatory design optimization}'', to give structure to the design process of AV, city, policy designers, and the public.
The workflow starts with PD activities, where stakeholders and citizens collaboratively identify (1) what metrics they desire to use to measure AV-city-policy design success and how they prioritize the metrics, and (2) what design decisions they want/need to participate in (rather than delegating to professionals). Next, using integrative simulation and Bayesian optimization, AV-city-policy designers can identify the designs that best reflect the stakeholders' priorities and preferred trade-offs. Finally, stakeholders and the public participate in the evaluation of the tentative, optimal designs before deployment.

By combining PD and quantitative optimization methods, this workflow frames the AV-city-policy design \textit{problem space }according to diverse stakeholders' agreed-upon assumptions and goals, while leveraging the efficiency of computational methods to search for optimal solutions within this problem space.

\begin{acks}
We thank the many researchers who have provided inputs to early drafts of this work.
This work is supported by the U.S. National Science Foundation under grants IIS-2107111 and IIS-2212431.
The last author is also partially supported by Schmidt Futures' AI2050 Early Career Fellowship.
\end{acks}

\bibliographystyle{ACM-Reference-Format}
\bibliography{paperpile,Sandhaus_manual}


\end{document}